\documentclass[useAMS,usenatbib]{mn2e}
\usepackage{graphicx}
\usepackage{amssymb,amsmath,amsfonts}
\usepackage{natbib}
\usepackage[section]{placeins}
\usepackage{verbatim}
\usepackage{enumitem}
\usepackage{multirow}
\usepackage{epsfig}
\usepackage{mathptmx}
\usepackage{booktabs,lscape,longtable}
\usepackage{graphicx}
\usepackage{comment}
\usepackage{color}
\usepackage[caption=false]{subfig}

\usepackage{url}

\title[The Brown dwarf Atmosphere Monitoring (BAM) Project II]{The Brown-dwarf Atmosphere Monitoring (BAM) Project II: Multi-epoch monitoring of extremely cool brown dwarfs}
\author[Rajan et al.]{A. Rajan$^{1}$\thanks{E-mail: arajan6@asu.edu}, J. Patience$^{1}$, P. A. Wilson$^{2}$, J. Bulger$^{1}$, R. J. De Rosa$^{1,2,3}$, K. Ward-Duong$^{1}$,\newauthor C. Morley$^{4}$, F. Pont$^{5}$, and R. Windhorst$^{1}$\\ 
$^{1}$School of Earth \& Space Exploration, 
   		Arizona State University, Tempe, AZ USA 85281\\
$^{2}$CNRS, Institut d'Astrophysique de Paris, UMR 7095,
        98bis boulevard Arago, 75014 Paris, France\\
$^{3}$University of California Berkeley,
        Berkeley, California, United States\\
$^{4}$Department of Astronomy and Astrophysics, 
        University of California, Santa Cruz, CA USA 95064\\
$^{5}$Astrophysics Group, School of Physics, 
		University of Exeter, Exeter EX4 4QL, UK}

\begin{document}

\date{Accepted XXXX. Received XXXX; in original form 2014 June 7}

\pagerange{\pageref{firstpage}--\pageref{lastpage}} \pubyear{2002}

\maketitle

\label{firstpage}

\begin{abstract}
With the discovery of Y dwarfs by the {\it WISE} mission, the population of field brown dwarfs now extends to objects with temperatures comparable to those of Solar System planets. To investigate the atmospheres of these newly identified brown dwarfs, we have conducted a pilot study monitoring an initial sample of three late T-dwarfs (T6.5, T8 and T8.5) and one Y-dwarf (Y0) for infrared photometric variability at multiple epochs. With $J$-band imaging, each target was observed for a period of 1.0~h to 4.5~h per epoch, which covers a significant fraction of the expected rotational period. These measurements represent the first photometric monitoring for these targets. For three of the four targets (2M1047, Ross~458C and WISE0458), multi-epoch monitoring was performed, with the time span between epochs ranging from a few hours to $\sim$2~years. During the first epoch, the T8.5 target WISE0458 exhibited variations with a remarkable min-to-max amplitude of 13~per~cent, while the second epoch light curve taken $\sim$2~years later did not note any variability to a 3~per~cent upper limit. With an effective temperature of $\sim$600 K, WISE0458 is the coldest variable brown dwarf published to-date, and combined with its high and variable amplitude makes it a fascinating target for detailed follow-up. The three remaining targets showed no significant variations, with a photometric precision between 0.8 and 20.0~per~cent, depending on the target brightness. Combining the new results with previous multi-epoch observations of brown dwarfs with spectral types of T5 or later, the currently identified variables have locations on the colour-colour diagram better matched by theoretical models incorporating cloud opacities rather than cloud-free atmospheres. This preliminary result requires further study to determine if there is a definitive link between variability among late-T dwarfs and their location on the colour-colour diagram.
\end{abstract}

\begin{keywords}
stars: variables: general, stars: low-mass, (stars:) brown dwarfs
\end{keywords}

\section{Introduction}
\begin{figure}
   \centering
	\includegraphics[width=8.5cm,trim=0.5cm 0.5cm 0.2cm 0.5cm]{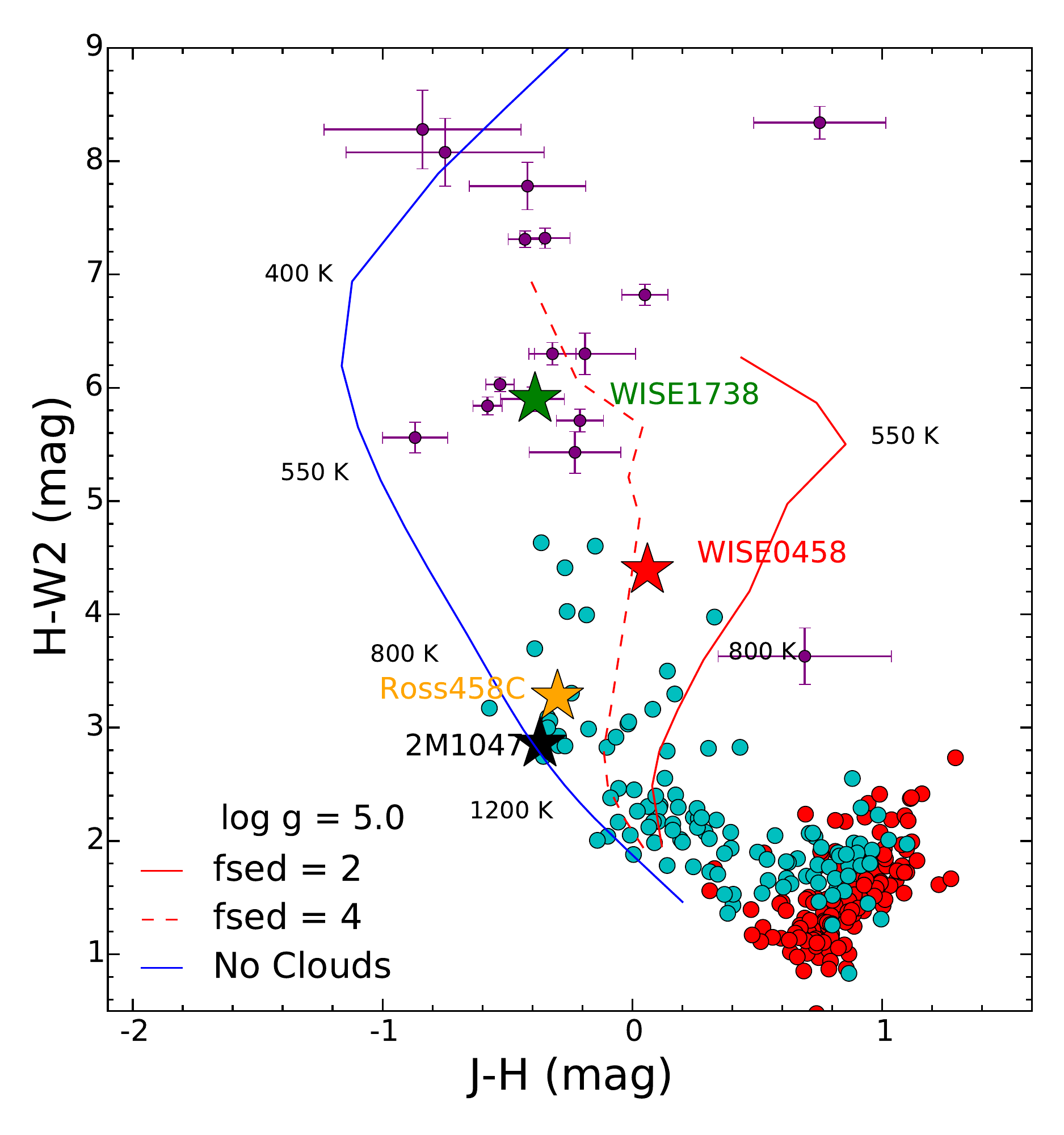}
    	\caption{ The L (red points), T (cyan points), and Y (purple points) brown dwarf sequence on an MKO $J, H$ and $W2$ ($WISE~4.6\micron$) colour-colour diagram. The blue solid line represents a cloudless atmosphere showing the change in photometric colour with temperature \citep{Saumon:2012}. The dashed red line and solid red line represent atmosphere models with an increasing cloud opacity from \citet{Morley:2012}. Magnitudes for the L, T \& Y dwarfs are from \citet{Dupuy:2012, Kirkpatrick:2012, Leggett:2015}. Details of the full sample are given in Table \ref{TargList}.
} \label{Fig:CMD}
   \end{figure}

Ultracool dwarfs, spanning the L, T, and recently discovered Y dwarf \citep{Kirkpatrick:1999, Cushing:2011} spectral types, provide a link between the coolest stars, giant planets in our Solar System, and exoplanets. Without sufficient mass for nuclear fusion \citep[e.g][]{Hayashi:1963}, brown dwarfs cool monotonically over time, causing changes in the chemical and physical processes responsible for sculpting the emergent spectra of their atmospheres. The formation and dissipation of dusty condensate clouds are key components of theoretical models developed to explain the fluxes and spectral features of brown dwarfs \citep[e.g][]{Allard:2001, Marley:2002, Burrows:2006, Helling:2006}. Early T dwarf atmospheric models predicted that once the clouds from the L/T transition sink below the photosphere, the subsequent T-sequence should remain cloud-free \citep{Marley:2002}. T-dwarfs however, appear to deviate from the expected cloud-free atmosphere colour as they cool (Figure~\ref{Fig:CMD}; blue curve) and become progressively redder. This phenomenon can be best explained by the formation of sulfide and alkali salt clouds as the brown dwarf cools \citep{Lodders:2006, Visscher:2006, Morley:2012}. With a three-dimensional treatment of the atmospheric dynamics, recent models have suggested that large-scale temperature variations may be present near the photospheres of brown dwarfs and that these regions with different temperatures may result in light curve variability \citep{Showman:2013, Robinson:2014}. \citet{Zhang:2014} also discuss how atmospheric turbulence and the resulting vortices and/or zonal jets can lead to photometric variability up to several percent over a wide range of timescales.

%

Photometric monitoring of brown dwarfs provides a method to search for evidence and evolution of cloud features, storms, or activity that cause surface brightness differences \citep[e.g.][]{Koen:2005,Artigau:2009,Buenzli:2014,Radigan:2014,BAM1}. The strength, persistence, and wavelength dependence of the variations reveal the underlying atmospheric processes of cloud formation, dissipation, and dynamics of the atmospheres \citep[e.g.][]{Showman:2013, Zhang:2014}. Over the past several years, we have conducted a large-scale photometric variability survey of brown dwarfs -– the Brown dwarf Atmosphere Monitoring (BAM) program. The initial BAM study \citep{BAM1} covered 69 targets spanning the L0-T8 sequence, and detected multiple late-T variables. In this paper, we present the results of the second component of the BAM project that is designed to search for photometric variability over several epochs in four ultracool brown dwarfs at or near the T/Y spectral boundary. The properties of the sample of T/Y dwarfs are summarised in Section 2. The near-infrared imaging observations are described in Section 3. In Section 4, the data reduction and analysis required to construct the light curves is detailed. Results for each target are given in Section 5, followed by the discussion in Section 6 that includes a comparison with samples of higher temperature L and T brown dwarfs and theoretical models of cool atmospheres.

\section{The BAM-II sample}
The sample for this pilot study consists of four ultracool field brown dwarfs which span the late-T to early-Y spectral types (see Figure \ref{Fig:CMD}). Table \ref{TargList} reports the target names, coordinates, $J$-band magnitudes,  spectral types, effective temperatures and distances. The targets were discovered as part of large-scale surveys with 2MASS \citep{Burgasser:1999}, UKIDSS \citep{Goldman:2010}, and \textsl{WISE} \citep{Cushing:2011}. For this initial study, the targets include objects with a range of properties, including youth \citep{West:2008}, radio emission \citep{Route:2012,Williams:2013}, and binarity \citep{Burgasser:2012}.

   \begin{table*}
   \centering
      \caption[]{Target list}
         \label{TargList}
         \begin{tabular}{lllccccc}
            \hline
            \noalign{\smallskip}
            \multicolumn{1}{c}{Target}  &  \multicolumn{1}{c}{RA}  &  \multicolumn{1}{c}{Dec}  &  J$_{\rm{MKO}}$ & SpT & Temperature & Distance & References\\
            & & & (mag) & & (K) & (pc) & \\
            \noalign{\smallskip}
            \hline
            \noalign{\smallskip}
            2MASSW~J1047539+212423 	& 10:47:53.8	& +21:24:23 & 15.46 &
            T6.5  & $\sim$900  & 10.6 & 1, 2, 3 \\
            
            Ross~458C			& 13:00:41.7	& +12:21:14
            & 16.71 & T8 & $\sim$650 & 11.7 & 4, 5, 6\\
            
            WISEP~J045853.89+643452.9AB	& 04:58:53.9	& +64:34:51.9	&
            17.47 & T8.5$^a$ & $\sim600^a$ & 14 & 7, 6\\
            
            WISEP~J173835.53+273258.9 	& 17:38:35.52	& +27:32:58.9	&
            20.05 & Y0 & 430$^{+50}_{-40}$  & 9.8  & 11, 9, 6\\	
            \noalign{\smallskip}
            \hline
         \end{tabular}
         
	$^a$ assigned value is for the binary and not the individual components.\\  
	\textbf{References}: [1] \citet{Burgasser:1999}, [2] \citet{Burgasser:2002}, [3] \citet{Vrba:2004}, [4] \citet{Goldman:2010}, [5] \citet{Burgasser:2010}, [6] \citet{Dupuy:2013}, [7] \citet{Mainzer:2010}, [8] \citet{Delorme:2010}, [9] \citet{Cushing:2011}, [10] \citet{Liu:2011} [11] \citet{Leggett:2013}.
			
   \end{table*}

The colour-colour diagram using near-infrared ($J, H$) and $WISE$ ($W2$) filters in Figure \ref{Fig:CMD} shows the variation in colours across the T and Y spectral types. Over-plotted is the clear atmosphere track from \citet{Saumon:2012} and theoretical models of brown dwarf atmospheres that include the effects of emergent sulfide clouds in the photosphere from \citet{Morley:2012}. Clouds with different values of the parameter $f_{\rm sed}$ have different cloud properties; low $f_{\rm sed}$ indicates smaller grain sizes and higher optical depth. Assuming that these clouds are patchy, we may expect to see more variability for the objects with colours consistent with more optically thick (low $f_{\rm sed}$) clouds. Our targets span a range of colours and model tracks, ranging from completely cloud-free (blue line), to low $f_{\rm sed}$ (solid red line) atmospheres with optically thick clouds, smaller grains, and larger vertical extent.

\subsection{2MASSW~J1047539+212423}

The brown dwarf 2MASSW~J1047539+212423 (2M1047) was identified with multi-epoch 2MASS images and classified as a T6.5 dwarf with Keck near-IR spectroscopy \citep{Burgasser:1999, Burgasser:2002}. Using a parallax-based distance of 10.3~pc an effective temperature of $\sim$900 K was inferred \citep{Vrba:2004}. 2M1047 is notable for measurement of radio variability, with bursts measured at a frequency of 4.75~GHz \citep{Route:2012}, and quasi-quiescent emission at 5.8~GHz \citep{Williams:2013}, making it the coolest brown dwarf with measured radio emission. Highly circularly polarised radio bursts lasting $\sim$100~s were detected in three of the fifteen observations with Arecibo each with a cadence of 0.1~s over $\sim$2~h. The radio observations were carried out over the course of one year, indicating a persistent source \citep{Route:2012}. The quasi-quiescent emission was of longer duration ($\sim$40 min), but also two orders of magnitude fainter than the radio bursts. The target was monitored for variability in the $J$ and $H$ bands by \citet{Artigau:2003} but did not exhibit any signs of variation. No contemporaneous optical or near-infrared photometric observations were recorded during either of the radio campaigns. 

\subsection{Ross~458C}

Ross~458C is the 102~arcsec common proper motion substellar companion to the stellar binary Ross~458AB \citep{Goldman:2010}, composed of two M-stars with a projected separation of $\sim$5~au \citep{Heintz:1994}. The companion was detected in the UKIRT Infrared Deep Sky Survey \citep[UKIDSS;][]{Lawrence:2007} and broadband/methane filter photometry identified Ross 458 C as a late-T dwarf \citep{Goldman:2010}. Subsequent near-infrared spectroscopy determined a spectral type of T8 and effective temperature of $\sim$650~K \citep{Burgasser:2010}. Similar assessments are reported in \citet{Burningham:2011} and \citet{Cushing:2011}. The distance to the Ross 458 system is $11.7^{+0.21}_{-0.20}$~pc \citep{Dupuy:2013}. The stellar pair in the system provides a means to estimate the age through measurements of stellar activity. The age for a field brown dwarf is hard to constrain. Based on the strength of H$\alpha$ emission, the level of variability \citep{West:2008} and space motion \citep{Montes:2001} of Ross~458AB, the age of the system has been estimated at $<$1~Gyr \citep[e.g.][]{Burgasser:2010, Burningham:2011}. Given the youth of the system, Ross~458C is predicted to be very low mass \citep[5-20~M$_{\rm{Jup}}$;][]{Burningham:2011}, which overlaps with planetary mass regime. Ross~458C is therefore a benchmark object for the investigation of both brown dwarf and exoplanet atmospheres. 

The atmosphere of Ross~458C measured by \citet{Burgasser:2010} with near-IR spectroscopy reveals evidence of low surface gravity and the authors were able to better fit the near-infrared spectrum with a cloudy atmosphere compared to cloudless. \citet{Burgasser:2010} initially proposed the clouds in the atmosphere to be the reemergence of iron and silicate clouds, but more recently \citet{Morley:2012} showed that models including salt and sulfide clouds fits the data better. Models incorporating both sulfide clouds and non-equilibrium chemistry have not yet been applied to the observational data.

\subsection{WISEP~J045853.89+643452.9AB}

WISEP~J045853.89+643452.9AB (WISE0458) was the first ultracool brown dwarf discovered by the Wide-field Infrared Survey Explorer satellite \citep[{\sl WISE};][]{Wright:2010} in its search for the coldest brown dwarfs in the solar neighborhood \citep{Mainzer:2010}. Comparison of the medium resolution near-infrared spectrum of WISE0458 with a grid of cloudless models suggested a very cool effective temperature of $\sim$600~K \citep{Mainzer:2010}. High angular resolution imaging with the Keck laser guide star AO system revealed that the system has a binary companion at a separation of $\sim$0\farcs5 and with a magnitude difference of $\sim$1~mag \citep{Gelino:2011}. Recent parallax measurements estimated the objects distance to be $14^{+5}_{-3}$~pc \citep{Dupuy:2013}. AO-assisted spatially resolved spectroscopy confirmed both objects are very late T dwarfs near the T/Y boundary; from the resolved spectra, the primary spectral type is T8.5 and the secondary spectral type is T9.5 \citep{Burgasser:2012}. We do not resolve the individual components in this study.

\subsection{WISEP~J173835.53+273258.9}

The final target, WISEP~J173835.53+273258.9 (WISE1738), is among the first Y dwarfs detected by the {\sl WISE} satellite \citep{Cushing:2011} and is still only one of 18 known Y dwarfs \citep{Kirkpatrick:2012,Leggett:2015}. With a spectral classification of Y0 and an effective temperature of 430$^{+50}_{-40}$~K \citep{Dupuy:2013}, this object is one of the coldest brown dwarfs discovered. WISE1738 has been selected as the spectral standard for the Y0 class \citep{Kirkpatrick:2012}. At these low temperatures water clouds are expected to form \citep{Burrows:2003}, potentially leading to variability in the emergent flux. Interestingly, WISE1738 might show long period photometric variability. The object was originally measured to have a $J$-band magnitude of $19.51\pm0.08$~mag in \citet{Kirkpatrick:2012}, which was subsequently measured to be $20.05\pm0.09$~mag in the recent \citet{Leggett:2013} study. We adopt the latter magnitude for the purposes of this study.  The \citet{Leggett:2013} study does note the inconsistency in the original \citet{Kirkpatrick:2012} photometry and suggests that the difference could be due to corrupted Palomar WIRC data. We adopt the \citet{Leggett:2013} magnitude for the purposes of this study. 


\section{Observations and Data Reduction}

This pilot study to monitor brown dwarfs at the T and Y dwarf boundary was initiated at the MMT observatory, with the first epoch of data for each of the four targets being obtained there. Follow-up observations were taken at three different observatories including the Canada-France-Hawaii Telescope (CFHT), UK Infrared Telescope (UKIRT), and the New Technology Telescope (NTT). The details for each of the targets is provided in the observing log presented in Table~\ref{tab:ObsLog}.

\subsection{MMT}

Observations of the entire target dataset were taken with the SAO Widefield InfraRed Camera \citep[SWIRC;][]{Brown:2008} at the 6.5~m MMT Observatory in Arizona, on the 12th and 13th of March 2012. The camera has an engineering grade $2048\times2048$ HAWAII-2 HgCdTe array, with a plate scale of 0.15~arcsec~pixel$^{-1}$, corresponding to an on-sky field-of-view of $5.12\times5.12$~arcmin. We employed the $J$-band filter on SWIRC, which closely matches the Mauna Kea Observatory (MKO) $J$-band filter \citep{Tokunaga:2005}. This filter was selected since the largest amplitude variations in known brown dwarf variables occur in the $J$-band \citep{Artigau:2009,Radigan:2012}. The observing strategy involved maintaining the target on a single pixel over an $\sim$20~min timescale with a four-point dither pattern for the purpose of sky subtraction; this sequence was repeated over a $\sim$1 -- 4 hour time period, as summarized in Table \ref{tab:ObsLog}. All of the raw images were calibrated using median combined darks and flat-field images. The exposure times ranged from 10s to 60s, depending on the target brightness, and the per-frame overhead for the detector was $\sim$5~s. The SWIRC detector has a scattered light artifact in the lower-left quadrant of the detector which was removed by using a high-pass filter. The SWIRC H2RG-detector is linear to within 0.1~per~cent up to 40000~DN and the exposure times were set to ensure that the target brown dwarfs were maintained well below the levels approaching the non-linear regime of the detector. Additionally, comparison stars with peak fluxes greater than 40000~DN were rejected from the analysis.

\subsection{CFHT}

We obtained $\sim$3~h of data on the T8.5 binary brown dwarf WISE0458 using the Wide-field InfraRed Camera (WIRCAM; \citealp{Puget:2004}) on the 3.5~m Canada France Hawaii Telescope (CFHT). With a plate scale of 0.3~arcsec~pixel$^{-1}$, corresponding to an on-sky field-of-view of $20\times20$~arcmin, there were several tens of similar-brightness reference stars within the field of each target for the differential photometry calculation. The observations were carried out in a queue-based observing mode in $J$-band with a median seeing of $\sim1\arcsec$ through most the sequence. The images were obtained in a staring mode, at a 60~s cadence. Per-frame overhead for the detector was $\sim$6.5~s. The data was dark and flat field calibrated using the `I`iwi automatic data pipeline from CFHT. The data was then sky subtracted in similar manner as described above, using a median filter (where the stars were first masked) to generate individual sky frames. The CFHT detector is linear to better than a~per~cent up to 5000~DN. The target and all the reference stars were exposed to less than this limit. Comparison stars brighter than 5000~DN were rejected in the analysis.


   \begin{table}
    \scriptsize{\centering
      \caption[]{Observing log}
         \label{tab:ObsLog}
         \begin{tabular}{lcccccccc}
            \hline
            \noalign{\smallskip}
            \centering Target & Telescope & Date & $\Delta$t & t$_{\rm exp}$ & Seeing$^{a}$\\
            \centering        &           &      &   (h)    & (s)           & (arcsec) \\
            \noalign{\smallskip}
            \hline
            \noalign{\smallskip}
            
            \multirow{3}{*}{2M1047}	& MMT & 2012-03-12T04:12:25 &  1.0 &  30 & 0.6 \\
                                    & MMT & 2012-03-12T09:37:52 &  1.9 &  30 & 0.8   \\
                                    & NTT & 2014-05-15T23:16:45 &  2.8 &  10 & 0.8   \\
                                    \hline
            \multirow{4}{*}{Ross~458C} & MMT   & 2012-03-12T07:34:24 & 1.5 &  30, 60 & 0.9 \\
                                       & UKIRT & 2014-04-23T07:02:54 & 2.5 &  10 & 0.9   \\
                                       & NTT   & 2014-05-16T23:42:38 & 5.3 & 20, 40 & 0.9\\
                                       & NTT   & 2014-05-20T04:18:56 & 5.1 & 40, 50, 70, 90 & 0.9\\
                                    \hline
            \multirow{2}{*}{WISE0458}  & MMT  & 2012-03-13T02:49:59 & 3.7  & 45 & 1.0 \\
                                       & CFHT & 2014-03-21T05:50:09 & 2.7  & 60 & 1.0 \\
                                    \hline
            WISE1738	& MMT & 2012-03-13T10:04:48 &  2.3 & 60 & 0.8 \\
            \noalign{\smallskip}
            \hline
         \end{tabular}}
        \\$^{a}$We report the median seeing for each target.
   \end{table}

\subsection{UKIRT}

The second epoch on Ross~458C was obtained on 23 April 2014 using the Wide-Field infrared Camera (WFCAM; \citealp{Casali:2007}), an infrared wide-field camera on the 3.8~m UK Infrared Telescope (UKIRT). WFCAM has four 2048x2048 array detectors with a pixel scale of $0.4$~arcsec~pixel$^{-1}$, corresponding to an on-sky field-of-view of $0.21\times0.21$~degrees. The observations were carried out with the MKO $J$-band filter, with seeing of $\sim0.9$~arcsec during the sequence. A five-point dither pattern was used, with each individual image having an exposure time of 10~s. Per-frame overhead for the detector was $\sim$1.3~s. The data were calibrated, and the different dithers were combined using the a dedicated pipeline developed by the Cambridge Astronomy Survey Unit \citep{Irwin:2008}. The WFCAM detectors are linear to better than a~per~cent within 40000 DN and care was taken to ensure that neither the target nor the reference stars used in the reduction were approaching this limit.

\subsection{NTT}

2M1047 and Ross~458C were observed between the 15th and 23rd of May 2014 with Son of ISAAC (SofI; \citealp{Moorwood:1998}) mounted on the 3.6~m NTT (New Technology Telescope). The observations utilized the wide-field imaging mode of SofI, with a plate scale of $0.28$~arcsec~pixel$^{-1}$, corresponding to an on-sky field-of-view of $4.92 \times 4.92$~arcmin. Per-frame overhead for the detector was $\sim$7.5~s. The NTT detector is linear to $>$1.5~per~cent when objects have less than 10000~DN peak flux. The observations were all obtained using the $J_{\rm s}$ filter, with a two-point AB-AB nodding pattern based on recommendations from the instrument scientist. The flux of the target, and reference stars within the field, was kept below 10,000~DN to minimise the effects from the detector non-linearity. To limit systematics in the data we used a two-point nod which permitted an accurate estimate of the sky background for the targets. For each object, data reduction consisting of correcting for the dark current and division by a flat field and sky subtraction were applied.

\section{Light Curve Generation and Identification of Variability}

We performed aperture photometry on the calibrated and aligned images from each of the observatories, using the \texttt{APPHOT} package in IRAF\footnote{IRAF is distributed by the National Optical Astronomy Observatories, which are operated by the Association of Universities for Research in Astronomy, Inc., under cooperative agreement with the National Science Foundation.}. For each of the targets in our study we computed the aperture photometry for a range of aperture sizes ranging from radii of 0.6 -- 2.0 times the full width half maximum (FWHM), and found that an aperture of 1.0~$\times$~FWHM provided the highest signal-to-noise ratio (SNR) across the full sample. Data taken during periods of poor conditions were not included in the analysis; the frame selection caused the gap between epoch 1 and 2 from 2M1047 and limited the W1738 observation period. A catalogue of all the stars in the field of view was generated, which typically included $10\sim50$ stars, from which the 15 reference stars most similar in brightness to the target were selected. We chose the similar brightness stars over the brightest stars for two reasons: to limit non-linearity effects, and to ensure that non-astrophysical variations due to weather were more accurately duplicated by the references. The limited field-of-view of the different detectors meant that there were not any objects of the same late-T spectral types to be used as references. 

   \begin{figure*}
   \centering
       \includegraphics[width=17.8cm,trim=0.5cm 0.5cm 0cm 0cm]{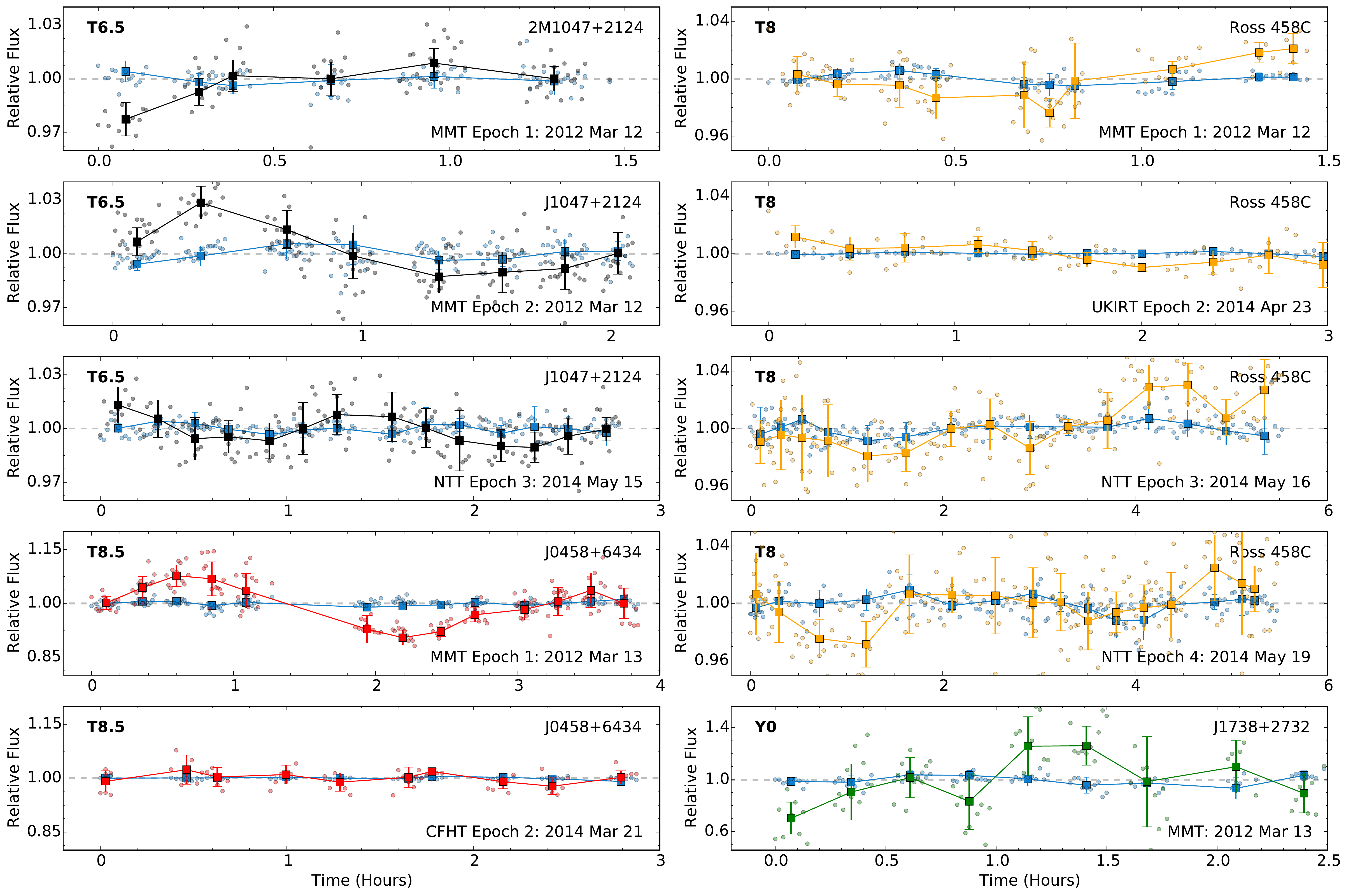}
       \caption{Detrended multi-epoch light curves for each of the targets with the master reference light curve (light blue). Top-left three panels are the different epochs for 2M1047 (black), with the date of observation and telescope noted in the panel. Bottom-left two panels are the same for W0458 (red). Top-right four panels are the multi-epoch observations for Ross~458C (orange). The bottom-right panel is the single epoch obtained for the Y0 brown dwarf WISE1738 (green).}
         \label{Fig:ConstLC}
   \end{figure*}
%

The final target light curves were generated by the photometric pipeline developed as part of the BAM-I survey \citep{BAM1}, modified to measure the photometry on each individual image. Measuring the photometry on individual images and taking the median value or forming a median image and measuring photometry on the median combined image did not result in significantly different light curves. The determination of whether the observed flux variation in the target light curve was due to an astrophysical process, required that the objects have a $p$-value~$\leq~5$~per~cent. The $p$-value is defined as the probability that the final target light curve is the same ($p$-value~$>$~10\%) or different ($p$-value~$\leq$~5\%) from the master reference light curve i.e. the median combination of all the reference light curves. Calculation of the $p$-value statistics for each of the light curves was carried out on weighted-mean combined data points where the errors are the 1~$\sigma$ scatter in each bin, rather than on the light curves composed of the individual photometry points. This was done to ensure that the statistics were not biased by outlier data points with small errors.

\section{Results}

The $J$-band light curves for each target generated using the observations at each epoch are shown in Figure~\ref{Fig:ConstLC}, plotted alongside the master reference light curve used to determine whether variations were observed. The master reference light curve is generated by median combining all the reference star light curves and shows any residual trends common to all light curves. A summary of the results of the light curve analysis, including the peak-to-peak amplitude of any variation and the associated $p$-value, is given in Table~\ref{tab:Results}. Each light curve is normalised, and variability is only investigated over the time scale of an individual epoch; no attempt to distinguish photometry variations between the individual epochs was made.

\begin{table*}
\centering
\caption{Summary of BAM-II variability study}
\label{tab:Results}
\begin{tabular}{lcccccccccc}
\hline
Object &   Spectral Type  & Telescope & Duration & Binned Images$^*$ & References & $\nu$ &  $\chi^2_\nu$ & $p$-value &   Amplitude/Limit$^{**}$ \\
 &  & &  (h) &  & & & & (\%) & (\%) \\
\hline
\hline
  \multirow{3}{*}{2M1047} &  \multirow{3}{*}{T6.5}  & MMT & 0.95 & 12 & 7 & 6      & 3      &   20 & $<0.8$\\
                          &                         & MMT & 1.88 & 19 & 7 & 10     & 1.92   &   10 & $<1.1$\\
                          &                         & NTT & 2.75 & 10 & 5 & 15     & 0.79   &   98 & $<1.0$\\
  
\hline
  
  \multirow{4}{*}{Ross~458C} &  \multirow{4}{*}{T8}  & MMT   & 1.78 & 7 & 6 & 11    & 0.58   &   10 & $<1.1$\\
                          &                          & UKIRT & 2.53 & 5 & 7 & 8     & 0.36   &   10 & $<0.8$\\
                          &                          & NTT & 5.29 & 10 & 9 & 13      & 0.41   &   94 & $<1.8$\\
                          &                          & NTT & 5.11 & 10 & 7 & 13      & 1.12   &   93 & $<2.1$\\
\hline
  
  \multirow{2}{*}{WISE0458} & \multirow{2}{*}{T8.5}  & MMT  & 3.68 & 11 & 8 & 12     & 11.37  &   0 & $13\pm3$\\
                          &                          & CFHT & 2.74 & 5 & 14 & 9     & 0.21   &   84 & $<2.6$\\
\hline
  WISE1738                &                     Y0   & MMT  & 2.39 & 8 & 10 & 5    & 0.9     &   15 & $<20.3$\\
\hline
\end{tabular}
\\Notes: $^*$The average number of individual images used to generate the weighted-mean light curves. $^{**}$ For the constant light curves the limit is the 1-$\sigma$ photometric error of the target light curve. For the variable epoch of W0458, the amplitude is the peak-to-trough value of the best fit sine curve to the data.
\end{table*}

\subsection{2M1047}
The time-series photometric measurements for the T6.5 dwarf 2M1047 are shown in the three top-left panels of Figure~\ref{Fig:ConstLC}. The first two epochs are separated by $\sim$3.5~h and the final epoch of data was taken 2.2~yr later. None of the three light curves for this target show statistically significant variability, and these results are similar to a previous $J$-band monitoring with the 1.6~m telescope of the L'observatoire du Mont-M\'{e}gantic by \citet{Artigau:2003}. Although there were no contemporaneous radio observations at the time of our $J$-band imaging, previous measurements of 2M1047 has shown variability at radio frequencies \citep[e.g.][]{Route:2012}. The radio bursts occurred over a time period of $\sim$100~s which is less than a single binned data point in the light curves. Radio bursts were recorded three times out of the fifteen observations (2~h each) made with Arecibo, but the current near-infrared data show no intensive brightening over comparable 1—3~h time intervals. In addition to the intense radio bursts, quasi-quiescent fluctuations in the radio emission with a timescale of $\sim$45~min have been reported \citep{Williams:2013}, however these variations were a factor of a hundred fainter than the large bursts described in \citet{Route:2012}. Contemporaneous radio and near-IR observations are required to search for any correlation between radio bursts and variations in photospheric flux.

\subsection{Ross~458C}
The four light curves for Ross~458C are given in the top-right panels of Figure~\ref{Fig:ConstLC}. The time spans between observations ranged from days to~yr, with 2.1~yr from the first to second epoch, 23.7~days between the second and third epoch, and 3.2~days between the third and fourth epoch. No statistically significant variations were detected at any of the four epochs, and the limits on detectable amplitudes ranged from 0.8 to 2.1~per~cent. The results suggest that there are no large and persistent storm features that would induce rotationally modulated brightness changes, or that the system is viewed pole-on. The Ross~458C data form the most comprehensive monitoring of the atmosphere of a brown dwarf that serves as an exoplanet analogue. Ross~458C represents the later stage of atmosphere evolution compared to the younger imaged exoplanets such as HR8799~d and $\beta$~Pic~b that may have similar masses but are substantially warmer due to their younger ages. Ross~458C also occupies an intermediate location in the colour-magnitude diagram between the youngest directly imaged planets and the older, cooler GJ504~b exoplanet \citep{Kuzuhara:2013}, but is far less technically challenging to monitor because of the wider angular separation between Ross~458AB and Ross~458C.

\subsection{WISE0458}
\begin{figure}
   \centering
       \includegraphics[width=8.5cm,trim=0.5cm 0.5cm 0cm 0cm]{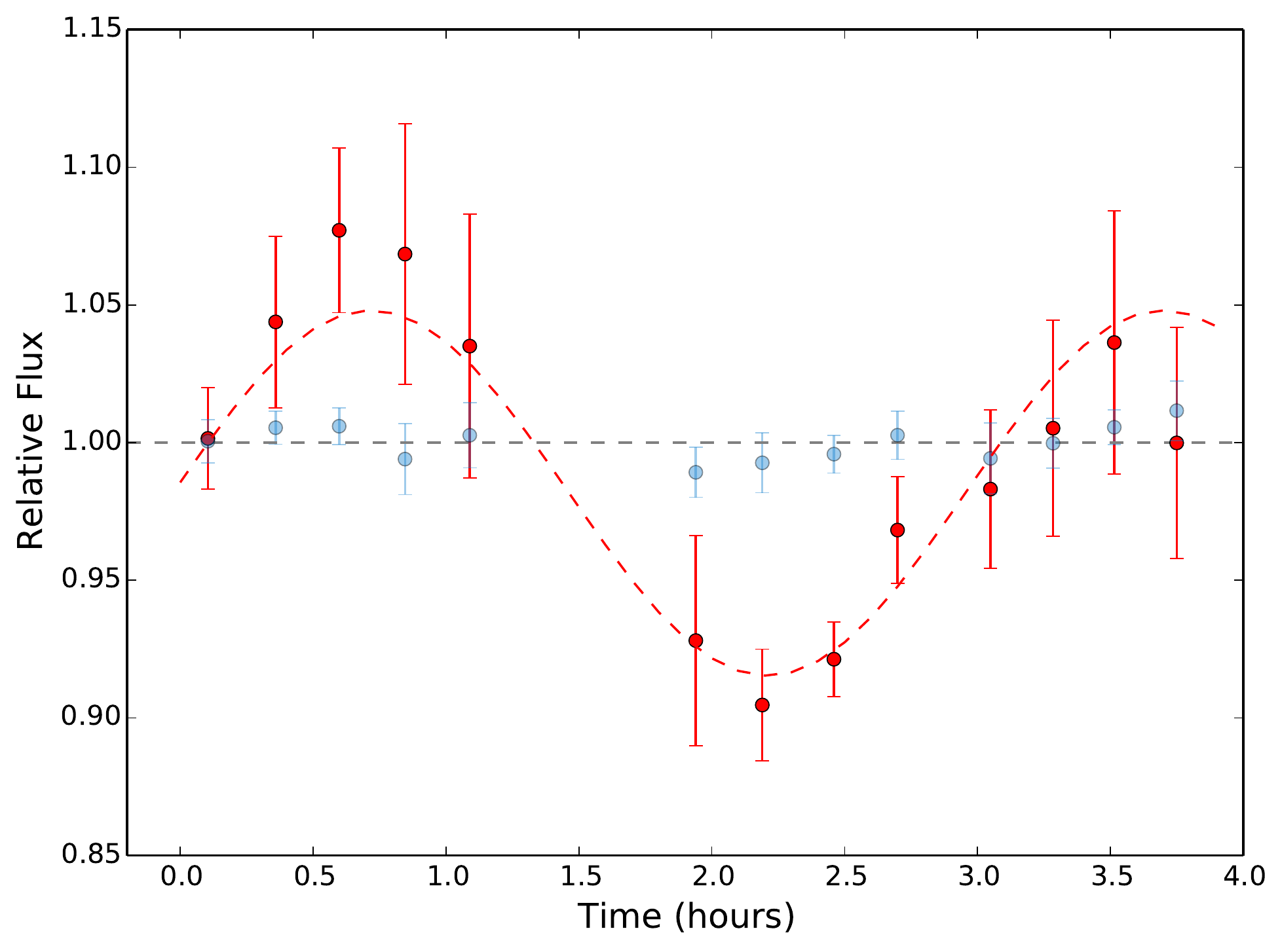}
       \caption{WISE0458 light curve (red circles) for the variable MMT epoch obtained on the 13th March 2012. Also plotted is the best fit sine wave (red dashed line) to the W0458 light curve resulting in a period of $\sim$3~h with a variable min-to-max amplitude of 13~per~cent. The master reference light curve is plotted with the light blue points.}
         \label{Fig:VarLC}
   \end{figure}

The bottom left two light curves in Figure~\ref{Fig:ConstLC} show the dramatic difference between the two epochs of observations for the binary brown dwarf WISE0458. In the first epoch, the target is highly variable compared to the master reference, with a min-to-max amplitude of $\sim17$~per~cent. The measured variability of WISE0458 is the second highest amplitude brown dwarf observed to date in a brown dwarf, with only 2M2139 exhibiting greater variability \citep{Radigan:2012}. The first epoch of WISE0458 data exhibits a periodic pattern and we fit a series of pure sine waves of different amplitudes and periods to the light curve. The best fit periodic signal with an amplitude of 13.2~per~cent and a period of 3~h is shown in Figure~\ref{Fig:VarLC}. The second epoch data on WISE0458 taken after a gap of $\sim$2~yr lacks any detectable variations. Further monitoring of WISE0458 will determine if the large amplitude variations recur.

%

\subsection{WISE1738}
The light curve for the Y0 target WISE1738, shown on the bottom right of Figure~\ref{Fig:ConstLC}, is impacted by the large uncertainties and the limited number of data points. During initial planning of the observing run at the MMT, WISE1738 was not expected to be the faintest target in our sample, however, as noted in Section 2, the photometry of this target was re-estimated to be 0.5 magnitudes fainter than originally measured \citep{Leggett:2013}. This intrinsic faintness combined with variable seeing during the observation resulted in low signal-to-noise ratio detections of the target and, consequently, an extremely noisy light curve. The light curve does not show any statistically significant variations, but the limitations in the photometric precision restrict the interpretation of the source as a constant with a large limit on possible variations of $<20$~per~cent.

\section{Discussion}

\begin{figure}
   \centering
	\includegraphics[width=9.5cm,trim=0.5cm 0.5cm 0.2cm 0.4cm]{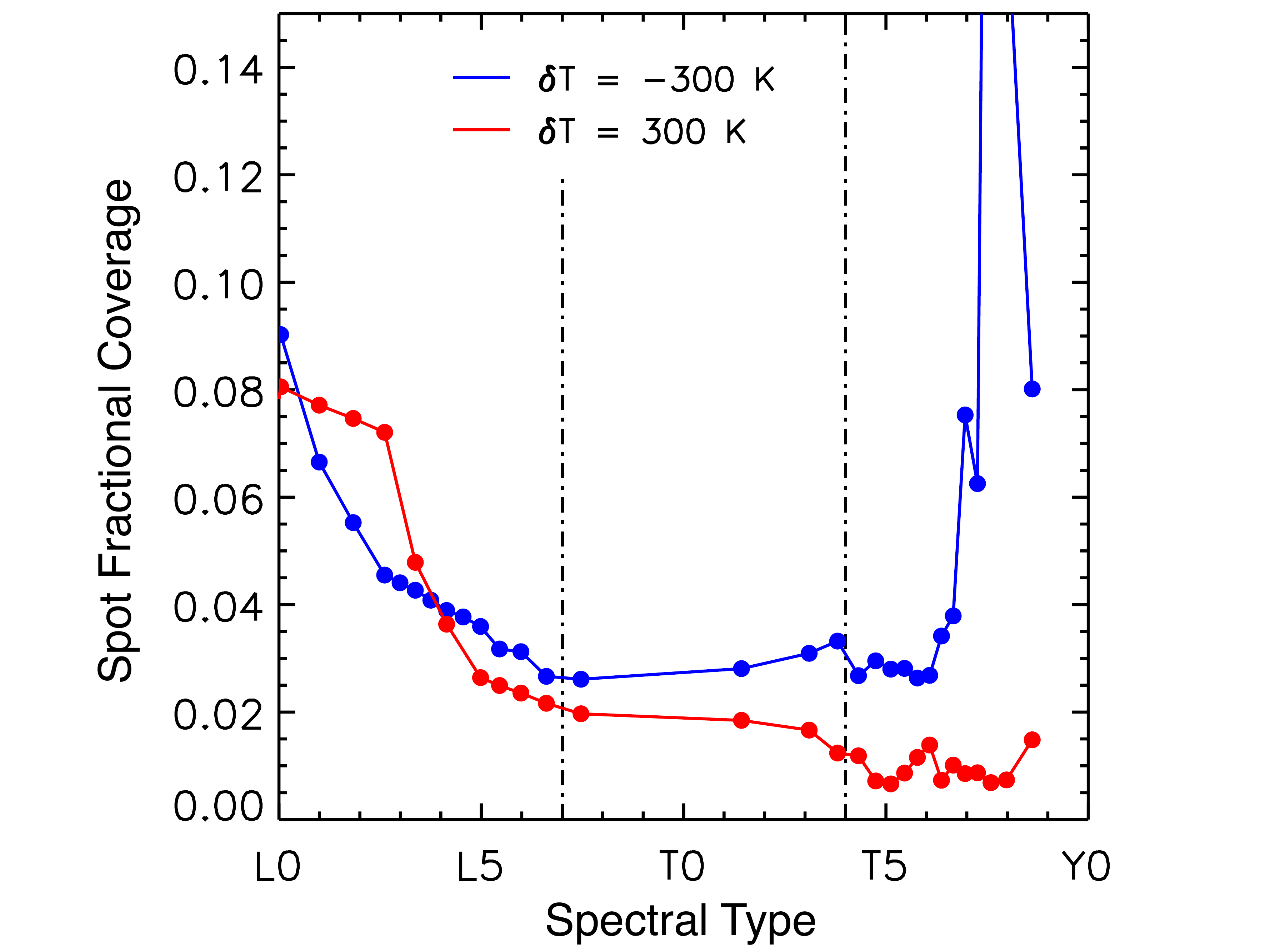}
    	\caption{Spot fractional coverage required for a 2~per~cent amplitude $J$-band light curve, plotted as a function of spectral type. The fractional coverage was estimated using the solar metallicity, $\log \left(g\right)=5.0$ synthetic spectra within the \textsc{BT-Settl} model grid \citep{Allard:2011}. For each $T_{\rm eff}$ within the grid, the fractional projected surface area at a higher/lower temperature $\left(\Delta T_{\rm eff}=\pm300\right)$ required to increase/decrease the $J$-band flux by 2~per~cent was estimated. This first-order approximation is analogous to the variability induced by a region of higher or lower temperature rotating in and out of view.}
    \label{Fig:SpotPlot}
\end{figure}
%
\begin{figure*}
   \centering
   \includegraphics[width=17.8cm,trim=0.5cm 0.5cm 0cm 0cm]{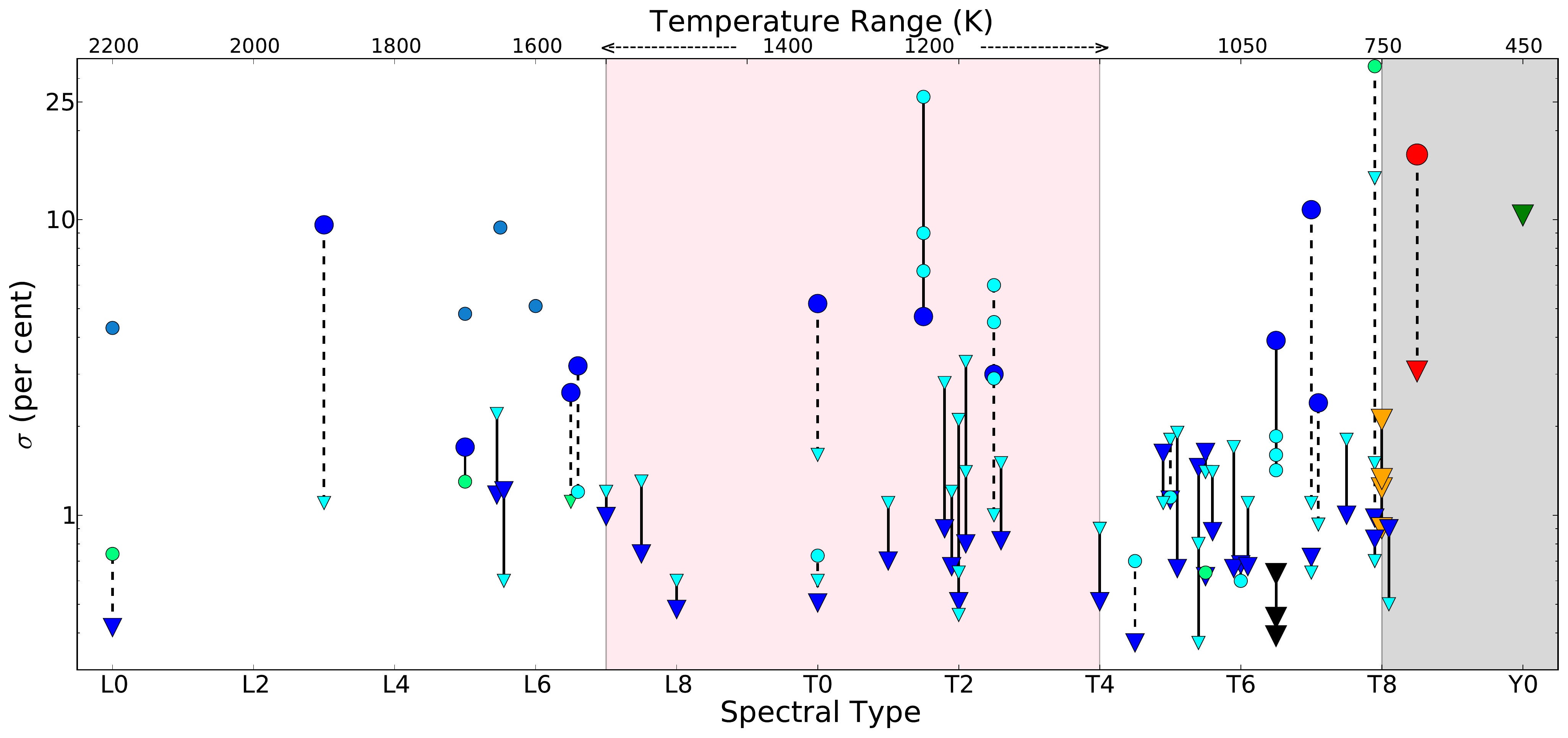}
   \caption{Plot summarizing literature multi-epoch variables and constants. Only objects with multiple epochs are plotted in the figure, with targets coming from \citep{Artigau:2009, Buenzli:2012, Clarke:2008, Girardin:2013, Khandrika:2013, Koen:2013, Metchev:2013, Radigan:2014, BAM1}. The BAM-II targets are indicated with black symbols for 2M1047, orange symbols for Ross~458C, red symbols for WISE0458 and green symbols for WISE1738. Each vertical line corresponds to a unique object, and solid lines indicate brown dwarfs that remained consistently variable or constant at more than one epoch, while the dashed lines identify the objects that switched between variable and constant states. The shaded regions indicate the L/T transition (pink) and T/Y boundary (gray) regions.}
         \label{Fig:surv_plot}
\end{figure*}
%

\begin{figure}
   \centering
	\includegraphics[width=8.5cm,trim=0.5cm 0.5cm 0.2cm 0.4cm]{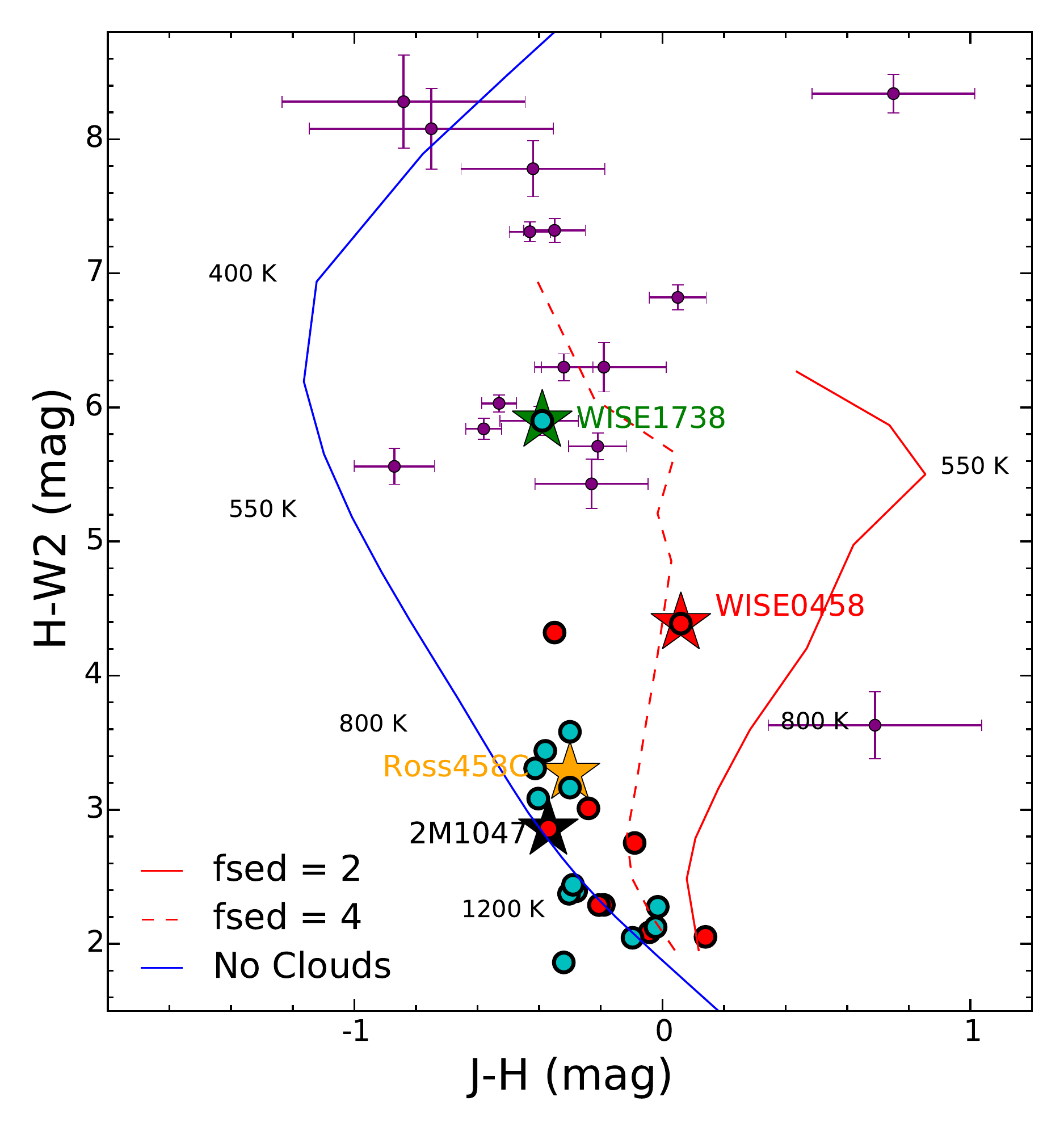}
    	\caption{ Colour-colour diagram showing the sample (large filled stars) and the Y dwarfs (purple) along with variable (red) and constant (blue) brown dwarfs with spectral type greater than T5 from Figure~\ref{Fig:surv_plot}. Figure~\ref{Fig:CMD} shows the full sample. The blue line is theoretical track for clear atmosphere from \citet{Saumon:2012} and the red lines indicating cloudy atmospheres are from \citet{Morley:2012}. All of the models assume a surface gravity of $\log$(g)~=~5.}
    \label{Fig:CCDv2}
\end{figure}
%

The BAM-II study was designed as a pilot program to investigate the coolest brown dwarf atmospheres using multi-epoch photometric monitoring as a probe of the dynamics and surface brightness variations. To place the BAM-II results in the broader context of the brown dwarf population, Figure~\ref{Fig:surv_plot} combines the targets in both this study, and the previous BAM-I survey spanning the full L0-T8 sequence \citep{BAM1}. Single epoch BAM-I variables are also included in the comparison. Additional epochs beyond the BAM-I measurements were provided by measurements from a recent large survey of L3-T9 brown dwarfs \citep{Radigan:2014} and from the more focused studies compiled in the BAM-I paper \citep{Artigau:2009, Buenzli:2012, Clarke:2008, Girardin:2013, Khandrika:2013, Koen:2013, Metchev:2013, Radigan:2012}. 

Nearly all of the amplitudes or limits shown in Figure~\ref{Fig:surv_plot} were measured in the $J$-band, however five objects include results from different filters, since this was the only way to include more than one epoch for those targets. Both the significant number of objects that switch between variable and constant states and the substantial range of amplitudes for the multi-epoch variables highlight the dynamic and evolving nature of substellar atmospheres, as well as the need for multi-epoch monitoring. Currently, the best studied case is that of SIMP0136 with monitoring from 2008 to 2012 revealing a remarkable evolution in both the amplitude (from $>5$~per~cent to undetectable) and the shape (from sinusoidal to multi-component) of the light curve measured over several hours per epoch \citep{Metchev:2013}. The BAM-I and BAM-II surveys have identified a set of targets spanning the full L-T sequence that warrant further monitoring.

The variable and constant brown dwarfs with spectral types of T5 and later that are included in Figure~\ref{Fig:surv_plot} are plotted on an ($H-W2$) vs. ($J-H$) colour-colour plot in Figure~\ref{Fig:CCDv2} to compare with a set of theoretical models (Saumon et al. 2012, Morley et al. 2012). A lack of {\it WISE} photometry for some targets in Figure~\ref{Fig:surv_plot} limits the number of brown dwarfs from this study that can be plotted in Figure~\ref{Fig:CCDv2}. The distribution of the late-T and Y dwarf population in this colour-colour diagram (larger sample in Figure~\ref{Fig:CMD}) cannot be reconciled with models of cloud-free atmospheres (see Figure~\ref{Fig:CCDv2}), and a proposed explanation for the dispersion involves varying amounts of opacity parametrised in the value of $f_{\rm sed}$ \citep{Morley:2012}. The redder $J-H$ colour of WISE0458 places the target close to the $f_{\rm sed}=4$ model with intermediate level of cloudiness as indicated in the colour-colour diagram of Figure~\ref{Fig:CCDv2}. The constant targets (blue circles) amongst the late-T objects with multi-epoch measurements presented in Figure~\ref{Fig:CCDv2}, appear concentrated near the cloud-free model, and the variables (red circles) appear to have redder and thus cloudier atmospheres. Changes in the $f_{\rm sed}$ values may be linked not only to the colour, but also to the presence or absence of variability. Determining whether or not there is a link between the variability and location in the colour-colour diagram requires a larger set of late-T monitoring observations.

Recent work on synthetic atmosphere models extending to the temperatures of late-T and Y dwarfs ($900-400$~K) has indicated that these brown dwarfs may have sulfide and alkali salt clouds in their photospheres (Morley et al. 2012). Such clouds peak in optical depth for objects with $T_{\rm eff}\sim 600$~K (T9) but persist in objects from 900~K to under 400~K. Three-dimensional models that include radiative transfer and cloud formation have not yet been developed for these cool objects. Idealised 3D-circulation simulations by \citet{Showman:2013} suggests that brown dwarfs may have complex circulation patterns on regional and global scales, generated by the interaction of convective layers with the overlying stably stratified radiative atmosphere, potentially leading to patchy cloud structure. As stated in the introduction, atmospheric circulation models including atmospheric turbulence predict a range of variability amplitudes over multiple epochs for the emitted flux \citet{Zhang:2014}. If cool brown dwarfs do have weather patterns causing heterogeneous cloud cover, the \citet{Morley:2012} models predict that they would show photometric variability in the near-infrared, predominantly in $Y$ and $J$-bands. Based on the initial results from this study, there is an indication that the variables have redder colours and are concentrated in the region of the colour-colour diagram associated with the atmosphere models more likely to have weather patterns. This suggestive link between theory and observations will be investigated further with larger surveys for multi-epoch variability among the coolest brown dwarfs.

Motivated by recent studies by \citet{Robinson:2014, Morley:2014}, we investigated what variability of different per~cent amplitudes mean in terms of actual hot/cold spot fractional coverage on the brown dwarf photosphere. We ran a simple simulation using the \textsc{BT-Settl} model grid from \citet{Allard:2011}, similar to what was done in \citet{Kostov:2013} but for a larger temperature range. In the simulation we estimated the spot coverage required to produce a particular amplitude of variability as a function of the spectral type, where the spectral types are defined by their temperature. Figure~\ref{Fig:SpotPlot} shows the results of a simulated 2~per~cent variable, for which the patches have a $\Delta$T$_{\rm eff}$ of $\pm$300K. The figure indicates that for different spectral types, the spot fractional coverage required to produce the same amplitude of variation can differ by several percent. And amplitude cutoffs to indicate ``strong'' or ``weak'' variations might not necessarily indicate higher and lower spot fractional coverage.

\section*{Acknowledgments}

Observations reported here were obtained at the MMT, CFHT, UKIRT, and NTT Observatories. The authors would like to extend their gratitude to the support staff at each observatory in enabling the successful implementation of the program. The authors also wish to recognize and acknowledge the very significant cultural role and reverence that the summit of Mauna Kea has always had within the indigenous Hawaiian community.  We are most fortunate to have the opportunity to conduct observations from this mountain. JP was supported by a Leverhulme research project grant (F/00144/BJ), and funding from an Science and Technology Facilities Council (STFC) standard grant. PAW acknowledges support from STFC. RJDR acknowledges support from STFC grants ST/H002707/1 and ST/K005588/1. This research has made use of the SIMBAD database and VizieR catalogue access tool, CDS, Strasbourg, France. This research has benefitted from the M, L, T, and Y dwarf compendium housed at DwarfArchives.org. This research made use of Astropy, a community-developed core Python package for Astronomy \citep{Astropy}.

\bibliographystyle{mn2e}
\setlength{\bibhang}{2.0em}
\setlength\labelwidth{0.0em}
\bibliography{myrefs}

\begin{thebibliography}{62}
\expandafter\ifx\csname natexlab\endcsname\relax\def\natexlab#1{#1}\fi

\bibitem[{{Allard} {et~al}\mbox{.}(2001){Allard}, {Hauschildt}, {Alexander},
  {Tamanai}, \& {Schweitzer}}]{Allard:2001}
{Allard} F., {Hauschildt} P.~H., {Alexander} D.~R., {Tamanai} A., {Schweitzer}
  A., 2001, \apj, 556, 357

\bibitem[{{Allard}, {Homeier} \& {Freytag}(2011){Allard}, {Homeier}, \&
  {Freytag}}]{Allard:2011}
{Allard} F., {Homeier} D., {Freytag} B., 2011, in Astronomical Society of the
  Pacific Conference Series, Vol. 448, 16th Cambridge Workshop on Cool Stars,
  Stellar Systems, and the Sun, {Johns-Krull} C., {Browning} M.~K., {West}
  A.~A., eds., p.~91

\bibitem[{{Artigau} {et~al}\mbox{.}(2009){Artigau}, {Bouchard}, {Doyon}, \&
  {Lafreni{\`e}re}}]{Artigau:2009}
{Artigau} {\'E}., {Bouchard} S., {Doyon} R., {Lafreni{\`e}re} D., 2009, \apj,
  701, 1534

\bibitem[{{Artigau}, {Nadeau} \& {Doyon}(2003){Artigau}, {Nadeau}, \&
  {Doyon}}]{Artigau:2003}
{Artigau} {\'E}., {Nadeau} D., {Doyon} R., 2003, in IAU Symposium, Vol. 211,
  Brown Dwarfs, {Mart{\'{\i}}n} E., ed., p. 451

\bibitem[{{Astropy Collaboration} {et~al}\mbox{.}(2013){Astropy Collaboration},
  {Robitaille}, {Tollerud}, {Greenfield}, {Droettboom}, {Bray}, {Aldcroft},
  {Davis}, {Ginsburg}, {Price-Whelan}, {Kerzendorf}, {Conley}, {Crighton},
  {Barbary}, {Muna}, {Ferguson}, {Grollier}, {Parikh}, {Nair}, {Unther},
  {Deil}, {Woillez}, {Conseil}, {Kramer}, {Turner}, {Singer}, {Fox}, {Weaver},
  {Zabalza}, {Edwards}, {Azalee Bostroem}, {Burke}, {Casey}, {Crawford},
  {Dencheva}, {Ely}, {Jenness}, {Labrie}, {Lim}, {Pierfederici}, {Pontzen},
  {Ptak}, {Refsdal}, {Servillat}, \& {Streicher}}]{Astropy}
{Astropy Collaboration} {et~al.}, 2013, \aap, 558, A33

\bibitem[{{Brown} {et~al}\mbox{.}(2008){Brown}, {McLeod}, {Geary}, \&
  {Bowsher}}]{Brown:2008}
{Brown} W.~R., {McLeod} B.~A., {Geary} J.~C., {Bowsher} E.~C., 2008, in SPIE
  Conference Series, Vol. 7014,

\bibitem[{{Buenzli} {et~al}\mbox{.}(2012){Buenzli}, {Apai}, {Morley},
  {Flateau}, {Showman}, {Burrows}, {Marley}, {Lewis}, \& {Reid}}]{Buenzli:2012}
{Buenzli} E. {et~al.}, 2012, \apjl, 760, L31

\bibitem[{{Buenzli} {et~al}\mbox{.}(2014){Buenzli}, {Apai}, {Radigan}, {Reid},
  \& {Flateau}}]{Buenzli:2014}
{Buenzli} E., {Apai} D., {Radigan} J., {Reid} I.~N., {Flateau} D., 2014, \apj,
  782, 77

\bibitem[{{Burgasser} {et~al}\mbox{.}(2012){Burgasser}, {Gelino}, {Cushing}, \&
  {Kirkpatrick}}]{Burgasser:2012}
{Burgasser} A.~J., {Gelino} C.~R., {Cushing} M.~C., {Kirkpatrick} J.~D., 2012,
  \apj, 745, 26

\bibitem[{{Burgasser} {et~al}\mbox{.}(2002){Burgasser}, {Kirkpatrick}, {Brown},
  {Reid}, {Burrows}, {Liebert}, {Matthews}, {Gizis}, {Dahn}, {Monet}, {Cutri},
  \& {Skrutskie}}]{Burgasser:2002}
{Burgasser} A.~J. {et~al.}, 2002, \apj, 564, 421

\bibitem[{{Burgasser} {et~al}\mbox{.}(1999){Burgasser}, {Kirkpatrick}, {Brown},
  {Reid}, {Gizis}, {Dahn}, {Monet}, {Beichman}, {Liebert}, {Cutri}, \&
  {Skrutskie}}]{Burgasser:1999}
{Burgasser} A.~J. {et~al.}, 1999, \apjl, 522, L65

\bibitem[{{Burgasser} {et~al}\mbox{.}(2010){Burgasser}, {Simcoe}, {Bochanski},
  {Saumon}, {Mamajek}, {Cushing}, {Marley}, {McMurtry}, {Pipher}, \&
  {Forrest}}]{Burgasser:2010}
{Burgasser} A.~J. {et~al.}, 2010, \apj, 725, 1405

\bibitem[{{Burningham} {et~al}\mbox{.}(2011){Burningham}, {Leggett}, {Homeier},
  {Saumon}, {Lucas}, {Pinfield}, {Tinney}, {Allard}, {Marley}, {Jones},
  {Murray}, {Ishii}, {Day-Jones}, {Gomes}, \& {Zhang}}]{Burningham:2011}
{Burningham} B. {et~al.}, 2011, \mnras, 414, 3590

\bibitem[{{Burrows}, {Sudarsky} \& {Hubeny}(2006){Burrows}, {Sudarsky}, \&
  {Hubeny}}]{Burrows:2006}
{Burrows} A., {Sudarsky} D., {Hubeny} I., 2006, \apj, 640, 1063

\bibitem[{{Burrows}, {Sudarsky} \& {Lunine}(2003){Burrows}, {Sudarsky}, \&
  {Lunine}}]{Burrows:2003}
{Burrows} A., {Sudarsky} D., {Lunine} J.~I., 2003, \apj, 596, 587

\bibitem[{{Casali} {et~al}\mbox{.}(2007){Casali}, {Adamson}, {Alves de
  Oliveira}, {Almaini}, {Burch}, {Chuter}, {Elliot}, {Folger}, {Foucaud},
  {Hambly}, {Hastie}, {Henry}, {Hirst}, {Irwin}, {Ives}, {Lawrence}, {Laidlaw},
  {Lee}, {Lewis}, {Lunney}, {McLay}, {Montgomery}, {Pickup}, {Read}, {Rees},
  {Robson}, {Sekiguchi}, {Vick}, {Warren}, \& {Woodward}}]{Casali:2007}
{Casali} M. {et~al.}, 2007, \aap, 467, 777

\bibitem[{{Clarke} {et~al}\mbox{.}(2008){Clarke}, {Hodgkin}, {Oppenheimer},
  {Robertson}, \& {Haubois}}]{Clarke:2008}
{Clarke} F.~J., {Hodgkin} S.~T., {Oppenheimer} B.~R., {Robertson} J., {Haubois}
  X., 2008, \mnras, 386, 2009

\bibitem[{{Cushing} {et~al}\mbox{.}(2011){Cushing}, {Kirkpatrick}, {Gelino},
  {Griffith}, {Skrutskie}, {Mainzer}, {Marsh}, {Beichman}, {Burgasser},
  {Prato}, {Simcoe}, {Marley}, {Saumon}, {Freedman}, {Eisenhardt}, \&
  {Wright}}]{Cushing:2011}
{Cushing} M.~C. {et~al.}, 2011, \apj, 743, 50

\bibitem[{{Delorme} {et~al}\mbox{.}(2010){Delorme}, {Albert}, {Forveille},
  {Artigau}, {Delfosse}, {Reyl{\'e}}, {Willott}, {Bertin}, {Wilkins}, {Allard},
  \& {Arzoumanian}}]{Delorme:2010}
{Delorme} P. {et~al.}, 2010, \aap, 518, A39

\bibitem[{{Dupuy} \& {Kraus}(2013)}]{Dupuy:2013}
{Dupuy} T.~J., {Kraus} A.~L., 2013, Science, 341, 1492

\bibitem[{{Dupuy} \& {Liu}(2012)}]{Dupuy:2012}
{Dupuy} T.~J., {Liu} M.~C., 2012, \apjs, 201, 19

\bibitem[{{Gelino} {et~al}\mbox{.}(2011){Gelino}, {Kirkpatrick}, {Cushing},
  {Eisenhardt}, {Griffith}, {Mainzer}, {Marsh}, {Skrutskie}, \&
  {Wright}}]{Gelino:2011}
{Gelino} C.~R. {et~al.}, 2011, \aj, 142, 57

\bibitem[{{Girardin}, {Artigau} \& {Doyon}(2013){Girardin}, {Artigau}, \&
  {Doyon}}]{Girardin:2013}
{Girardin} F., {Artigau} {\'E}., {Doyon} R., 2013, \apj, 767, 61

\bibitem[{{Goldman} {et~al}\mbox{.}(2010){Goldman}, {Marsat}, {Henning},
  {Clemens}, \& {Greiner}}]{Goldman:2010}
{Goldman} B., {Marsat} S., {Henning} T., {Clemens} C., {Greiner} J., 2010,
  \mnras, 405, 1140

\bibitem[{{Hayashi} \& {Nakano}(1963)}]{Hayashi:1963}
{Hayashi} C., {Nakano} T., 1963, Progress of Theoretical Physics, 30, 460

\bibitem[{{Heintz}(1994)}]{Heintz:1994}
{Heintz} W.~D., 1994, \aj, 108, 2338

\bibitem[{{Helling} \& {Woitke}(2006)}]{Helling:2006}
{Helling} C., {Woitke} P., 2006, \aap, 455, 325

\bibitem[{{Irwin}(2008)}]{Irwin:2008}
{Irwin} M.~J., 2008, in 2007 ESO Instrument Calibration Workshop, {Kaufer} A.,
  {Kerber} F., eds., p. 541

\bibitem[{{Khandrika} {et~al}\mbox{.}(2013){Khandrika}, {Burgasser}, {Melis},
  {Luk}, {Bowsher}, \& {Swift}}]{Khandrika:2013}
{Khandrika} H., {Burgasser} A.~J., {Melis} C., {Luk} C., {Bowsher} E., {Swift}
  B., 2013, \aj, 145, 71

\bibitem[{{Kirkpatrick} {et~al}\mbox{.}(2012){Kirkpatrick}, {Gelino},
  {Cushing}, {Mace}, {Griffith}, {Skrutskie}, {Marsh}, {Wright}, {Eisenhardt},
  {McLean}, {Mainzer}, {Burgasser}, {Tinney}, {Parker}, \&
  {Salter}}]{Kirkpatrick:2012}
{Kirkpatrick} J.~D. {et~al.}, 2012, \apj, 753, 156

\bibitem[{{Kirkpatrick} {et~al}\mbox{.}(1999){Kirkpatrick}, {Reid}, {Liebert},
  {Cutri}, {Nelson}, {Beichman}, {Dahn}, {Monet}, {Gizis}, \&
  {Skrutskie}}]{Kirkpatrick:1999}
{Kirkpatrick} J.~D. {et~al.}, 1999, \apj, 519, 802

\bibitem[{{Koen}(2013)}]{Koen:2013}
{Koen} C., 2013, \mnras, 428, 2824

\bibitem[{{Koen} {et~al}\mbox{.}(2005){Koen}, {Tanab{\'e}}, {Tamura}, \&
  {Kusakabe}}]{Koen:2005}
{Koen} C., {Tanab{\'e}} T., {Tamura} M., {Kusakabe} N., 2005, \mnras, 362, 727

\bibitem[{{Kostov} \& {Apai}(2013)}]{Kostov:2013}
{Kostov} V., {Apai} D., 2013, \apj, 762, 47

\bibitem[{{Kuzuhara} {et~al}\mbox{.}(2013){Kuzuhara}, {Tamura}, {Kudo},
  {Janson}, {Kandori}, {Brandt}, {Thalmann}, {Spiegel}, {Biller}, {Carson},
  {Hori}, {Suzuki}, {Burrows}, {Henning}, {Turner}, {McElwain},
  {Moro-Mart{\'{\i}}n}, {Suenaga}, {Takahashi}, {Kwon}, {Lucas}, {Abe},
  {Brandner}, {Egner}, {Feldt}, {Fujiwara}, {Goto}, {Grady}, {Guyon},
  {Hashimoto}, {Hayano}, {Hayashi}, {Hayashi}, {Hodapp}, {Ishii}, {Iye},
  {Knapp}, {Matsuo}, {Mayama}, {Miyama}, {Morino}, {Nishikawa}, {Nishimura},
  {Kotani}, {Kusakabe}, {Pyo}, {Serabyn}, {Suto}, {Takami}, {Takato}, {Terada},
  {Tomono}, {Watanabe}, {Wisniewski}, {Yamada}, {Takami}, \&
  {Usuda}}]{Kuzuhara:2013}
{Kuzuhara} M. {et~al.}, 2013, \apj, 774, 11

\bibitem[{{Lawrence} {et~al}\mbox{.}(2007){Lawrence}, {Warren}, {Almaini},
  {Edge}, {Hambly}, {Jameson}, {Lucas}, {Casali}, {Adamson}, {Dye}, {Emerson},
  {Foucaud}, {Hewett}, {Hirst}, {Hodgkin}, {Irwin}, {Lodieu}, {McMahon},
  {Simpson}, {Smail}, {Mortlock}, \& {Folger}}]{Lawrence:2007}
{Lawrence} A. {et~al.}, 2007, \mnras, 379, 1599

\bibitem[{{Leggett} {et~al}\mbox{.}(2015){Leggett}, {Morley}, {Marley}, \&
  {Saumon}}]{Leggett:2015}
{Leggett} S.~K., {Morley} C.~V., {Marley} M.~S., {Saumon} D., 2015, \apj, 799,
  37

\bibitem[{{Leggett} {et~al}\mbox{.}(2013){Leggett}, {Morley}, {Marley},
  {Saumon}, {Fortney}, \& {Visscher}}]{Leggett:2013}
{Leggett} S.~K., {Morley} C.~V., {Marley} M.~S., {Saumon} D., {Fortney} J.~J.,
  {Visscher} C., 2013, \apj, 763, 130

\bibitem[{{Liu} {et~al}\mbox{.}(2011){Liu}, {Delorme}, {Dupuy}, {Bowler},
  {Albert}, {Artigau}, {Reyl{\'e}}, {Forveille}, \& {Delfosse}}]{Liu:2011}
{Liu} M.~C. {et~al.}, 2011, \apj, 740, 108

\bibitem[{{Lodders} \& {Fegley}(2006)}]{Lodders:2006}
{Lodders} K., {Fegley}, Jr. B., 2006, {Chemistry of Low Mass Substellar
  Objects}, {Mason} J.~W., ed., p.~1

\bibitem[{{Mainzer} {et~al}\mbox{.}(2011){Mainzer}, {Cushing}, {Skrutskie},
  {Gelino}, {Kirkpatrick}, {Jarrett}, {Masci}, {Marley}, {Saumon}, {Wright},
  {Beaton}, {Dietrich}, {Eisenhardt}, {Garnavich}, {Kuhn}, {Leisawitz},
  {Marsh}, {McLean}, {Padgett}, \& {Rueff}}]{Mainzer:2010}
{Mainzer} A. {et~al.}, 2011, \apj, 726, 30

\bibitem[{{Marley} {et~al}\mbox{.}(2002){Marley}, {Seager}, {Saumon},
  {Lodders}, {Ackerman}, {Freedman}, \& {Fan}}]{Marley:2002}
{Marley} M.~S., {Seager} S., {Saumon} D., {Lodders} K., {Ackerman} A.~S.,
  {Freedman} R.~S., {Fan} X., 2002, \apj, 568, 335

\bibitem[{{Metchev} {et~al}\mbox{.}(2013){Metchev}, {Apai}, {Radigan},
  {Artigau}, {Heinze}, {Helling}, {Homeier}, {Littlefair}, {Morley}, {Skemer},
  \& {Stark}}]{Metchev:2013}
{Metchev} S. {et~al.}, 2013, Astronomische Nachrichten, 334, 40

\bibitem[{{Montes} {et~al}\mbox{.}(2001){Montes}, {L{\'o}pez-Santiago},
  {G{\'a}lvez}, {Fern{\'a}ndez-Figueroa}, {De Castro}, \&
  {Cornide}}]{Montes:2001}
{Montes} D., {L{\'o}pez-Santiago} J., {G{\'a}lvez} M.~C.,
  {Fern{\'a}ndez-Figueroa} M.~J., {De Castro} E., {Cornide} M., 2001, \mnras,
  328, 45

\bibitem[{{Moorwood}, {Cuby} \& {Lidman}(1998){Moorwood}, {Cuby}, \&
  {Lidman}}]{Moorwood:1998}
{Moorwood} A., {Cuby} J.-G., {Lidman} C., 1998, The Messenger, 91, 9

\bibitem[{{Morley} {et~al}\mbox{.}(2012){Morley}, {Fortney}, {Marley},
  {Visscher}, {Saumon}, \& {Leggett}}]{Morley:2012}
{Morley} C.~V., {Fortney} J.~J., {Marley} M.~S., {Visscher} C., {Saumon} D.,
  {Leggett} S.~K., 2012, \apj, 756, 172

\bibitem[{{Morley} {et~al}\mbox{.}(2014){Morley}, {Marley}, {Fortney}, \&
  {Lupu}}]{Morley:2014}
{Morley} C.~V., {Marley} M.~S., {Fortney} J.~J., {Lupu} R., 2014, \apjl, 789,
  L14

\bibitem[{{Puget} {et~al}\mbox{.}(2004){Puget}, {Stadler}, {Doyon}, {Gigan},
  {Thibault}, {Luppino}, {Barrick}, {Benedict}, {Forveille}, {Rambold},
  {Thomas}, {Vermeulen}, {Ward}, {Beuzit}, {Feautrier}, {Magnard}, {Mella},
  {Preis}, {Vallee}, {Wang}, {Lin}, {Hall}, \& {Hodapp}}]{Puget:2004}
{Puget} P. {et~al.}, 2004, in Society of Photo-Optical Instrumentation
  Engineers (SPIE) Conference Series, Vol. 5492, Ground-based Instrumentation
  for Astronomy, {Moorwood} A.~F.~M., {Iye} M., eds., pp. 978--987

\bibitem[{{Radigan} {et~al}\mbox{.}(2012){Radigan}, {Jayawardhana},
  {Lafreni{\`e}re}, {Artigau}, {Marley}, \& {Saumon}}]{Radigan:2012}
{Radigan} J., {Jayawardhana} R., {Lafreni{\`e}re} D., {Artigau} {\'E}.,
  {Marley} M., {Saumon} D., 2012, \apj, 750, 105

\bibitem[{{Radigan} {et~al}\mbox{.}(2014){Radigan}, {Lafreni{\`e}re},
  {Jayawardhana}, \& {Artigau}}]{Radigan:2014}
{Radigan} J., {Lafreni{\`e}re} D., {Jayawardhana} R., {Artigau} E., 2014, \apj,
  793, 75

\bibitem[{{Robinson} \& {Marley}(2014)}]{Robinson:2014}
{Robinson} T.~D., {Marley} M.~S., 2014, \apj, 785, 158

\bibitem[{{Route} \& {Wolszczan}(2012)}]{Route:2012}
{Route} M., {Wolszczan} A., 2012, \apjl, 747, L22

\bibitem[{{Saumon} {et~al}\mbox{.}(2012){Saumon}, {Marley}, {Abel},
  {Frommhold}, \& {Freedman}}]{Saumon:2012}
{Saumon} D., {Marley} M.~S., {Abel} M., {Frommhold} L., {Freedman} R.~S., 2012,
  \apj, 750, 74

\bibitem[{{Showman} \& {Kaspi}(2013)}]{Showman:2013}
{Showman} A.~P., {Kaspi} Y., 2013, \apj, 776, 85

\bibitem[{{Tokunaga} \& {Vacca}(2005)}]{Tokunaga:2005}
{Tokunaga} A.~T., {Vacca} W.~D., 2005, \pasp, 117, 421

\bibitem[{{Visscher}, {Lodders} \& {Fegley}(2006){Visscher}, {Lodders}, \&
  {Fegley}}]{Visscher:2006}
{Visscher} C., {Lodders} K., {Fegley}, Jr. B., 2006, \apj, 648, 1181

\bibitem[{{Vrba} {et~al}\mbox{.}(2004){Vrba}, {Henden}, {Luginbuhl}, {Guetter},
  {Munn}, {Canzian}, {Burgasser}, {Kirkpatrick}, {Fan}, {Geballe},
  {Golimowski}, {Knapp}, {Leggett}, {Schneider}, \& {Brinkmann}}]{Vrba:2004}
{Vrba} F.~J. {et~al.}, 2004, \aj, 127, 2948

\bibitem[{{West} {et~al}\mbox{.}(2008){West}, {Hawley}, {Bochanski}, {Covey},
  {Reid}, {Dhital}, {Hilton}, \& {Masuda}}]{West:2008}
{West} A.~A., {Hawley} S.~L., {Bochanski} J.~J., {Covey} K.~R., {Reid} I.~N.,
  {Dhital} S., {Hilton} E.~J., {Masuda} M., 2008, \aj, 135, 785

\bibitem[{{Williams}, {Berger} \& {Zauderer}(2013){Williams}, {Berger}, \&
  {Zauderer}}]{Williams:2013}
{Williams} P.~K.~G., {Berger} E., {Zauderer} B.~A., 2013, \apjl, 767, L30

\bibitem[{{Wilson}, {Rajan} \& {Patience}(2014){Wilson}, {Rajan}, \&
  {Patience}}]{BAM1}
{Wilson} P.~A., {Rajan} A., {Patience} J., 2014, \aap, 566, A111

\bibitem[{{Wright} {et~al}\mbox{.}(2010){Wright}, {Eisenhardt}, {Mainzer},
  {Ressler}, {Cutri}, {Jarrett}, {Kirkpatrick}, {Padgett}, {McMillan},
  {Skrutskie}, {Stanford}, {Cohen}, {Walker}, {Mather}, {Leisawitz}, {Gautier},
  {McLean}, {Benford}, {Lonsdale}, {Blain}, {Mendez}, {Irace}, {Duval}, {Liu},
  {Royer}, {Heinrichsen}, {Howard}, {Shannon}, {Kendall}, {Walsh}, {Larsen},
  {Cardon}, {Schick}, {Schwalm}, {Abid}, {Fabinsky}, {Naes}, \&
  {Tsai}}]{Wright:2010}
{Wright} E.~L. {et~al.}, 2010, \aj, 140, 1868

\bibitem[{{Zhang} \& {Showman}(2014)}]{Zhang:2014}
{Zhang} X., {Showman} A.~P., 2014, \apjl, 788, L6

\end{thebibliography}

\label{lastpage}

\end{document}